\documentclass{article}
\usepackage{amsfonts,amssymb, amsmath}
\usepackage[english]{babel}
\usepackage{hyperref}

\newtheorem{prop}{Proposition}

\newtheorem{ex}{Example}

\def\bq{ \begin{equation}}
\def\eq{ \end{equation}}
\def\ben{ \begin{eqnarray}}
\def\en{ \end{eqnarray}}

\def\one#1{#1^{\raise5pt\hbox{$\scriptstyle\!\!\!\!1$}}\,{}}
\def\two#1{#1^{\raise5pt\hbox{$\scriptstyle\!\!\!\!2$}}\,{}}

\begin{document}


\title{Reduction of divisors for classical superintegrable $GL(3)$ magnetic chain}

\author{A.V. Tsiganov \\
\it\small St. Petersburg State University, St. Petersburg, Russia\\
\it\small e--mail: andrey.tsiganov@gmail.com}
\date{}
\maketitle

\begin{abstract}
 Variables of separation for classical $GL(3)$ magnetic chain obtained by Sklyanin form a generic positive divisor $D$ of degree $n$ on a genus $g$ non-hyperelliptic algebraic curve. Because $n>g$ this divisor $D$ has unique representative $\rho(D)$ in the Jacobian which can be constructed by using dim$|D|=n-g$ steps of Abel's algorithm. We study properties of the corresponding chain of divisors and prove that the classical $GL(3)$ magnetic chain is a superintegrable system with dim$|D|=2$ superintegrable Hamiltonians.
 \end{abstract}
\maketitle

\section{Introduction}
\setcounter{equation}{0}
Following to Nekhoroshev \cite{neh72}, integrable system is called  degenerate integrable or superintegrable system if the corresponding Hamiltonian vector field on a $2n$-dimensional symplectic manifold
\[
X=\Pi\mathrm{d}H
\]
 has $n+k$ independent first integrals such that only $n$ of them are in involution to each other
\[
 \{h_i,h_j\}=0\,,\qquad i,j=1,\ldots, n.
\]
 Here $H$ is superintegrable Hamiltonian and $\Pi$ is the Poisson bivector associated with the Poisson bracket $\{.,.\}$. Without loss of generality we can always take $H=h_1$ and suppose that remaining $k$ first integrals $u_1,\ldots,u_k$ do not commute with other $n-1$ first integrals $h_2,\ldots,h_n$.

In this note we study degenerate integrable or superintegrable systems with $k$ first integrals $u_1,\ldots,u_k$, which commute not only with superintegrable Hamiltonian $H=h_1$ but also with  a few  superintegrable Hamiltonians $H_2=h_2,\ldots,H_\ell=h_\ell$. According \cite{ts19ar,ts20}  these superintegrable systems are closely related with the Abel and Riemann-Roch theorems.

Let $C$ be an irreducible algebraic curve of genus $g$, canonically embedded in $\mathbb P^{g-1}$. In terms of points on $C$
the Riemann-Roch theorem looks like
\begin{equation}\label{rr-t}
\mathrm{dim}|D|=\mathrm{deg}(D) - 1 - \mathrm{dim}\, span(D),
\end{equation}
or, equivalently,
\[
\mathrm{dim}\, L(D)=\mathrm{deg}(D)  - \mathrm{dim}\, span(D)
\]
Here $D=P_1+P_2+\cdots+P_n$ is a positive divisor, i.e. formal sum of the points $P_i=(x_i,y_i)$ on a plane curve $C$ where may be repetition among the points $P_i$. The complete linear system $|D|$ is the set of all nonnegative divisors which are linearly equivalent to $D$
\[
|D|=\{D'\in \mathrm{Div}(C)\,|\,D'\sim D\,\mathrm{and}\,D'>0\}\,.
\]
The $span(D)$ is the intersection of all the hyperplanes in $\mathbb P^{g-1}$ containing $D$ and $L(D)$ is the space of meromorphic functions with poles bounded by $D$. For a general positive divisor $D$ of degree $n\geq g$ equation (\ref{rr-t}) has the following form
\begin{equation}\label{rr-t-red}
\mathrm{dim} |D|=n-g\,,
\end{equation}
see details in textbook \cite{mir}.

In classical mechanics positive divisors without matching points
\begin{equation}\label{eff-d}
D=P_1+\ldots+P_n\,,\qquad \mathrm{deg}(D)=n
\end{equation}
appear in a study of dynamical systems on the $2n$-dimensional smooth symplectic manifold integrable by Abel's quadratures
\begin{equation}\label{ab-eq}
\sum_{i=1}^n\int^{P_i}\omega_j=t_j\,,\qquad j=1,\ldots,n,
\end{equation}
Here $\omega_j$ are independent differentials on a genus $g$ algebraic  curve $C$ defined by the equation
\[
C:\qquad f(x,y)=0\,.
\]
This equation  on a projective plane appears either as a characteristic equation associated with the Lax matrix $L(x)$ with spectral parameter
\[
f(x,y)=\mathrm{det}(L(x)-y)=0
\]
 or as separation relation in the Jacobi separation of variables method
\[
f(x_i,y_i,,h_1,\ldots,h_n)=0
\]
binding together $n$ pairs of variables of separation $(x_i, y_i)$ and independent first integrals $h_1,\ldots, h_n$ in involution
\[
\{h_i,h_j\}=0\,,\qquad i,j=1,\ldots,n,
\]
see  \cite{bt03,skl95} and references within.

Let us suppose  that  Abel's quadratures (\ref{ab-eq}) consist of $g$ quadratures
\begin{equation}\label{ab-eq-g}
\sum_{i=1}^n\int^{P_i}\omega_j=t_j\,,\qquad j=1,\ldots,g\,,\qquad t_j\in \mathbb C\,,
\end{equation}
with independent differentials of the first kind $\omega_1,\ldots, \omega_g$ (everywhere-regular differential 1-forms)
and dim$|D|=n-g$ quadratures with independent non-regular differentials, meromorphic differentials, differentials of the second or  third kind, etc. According to Abel's theorem \cite{ab} subsystem of Abel's equations (\ref{ab-eq-g}) has $k=g$  integrals $u_1,\ldots,u_{g}$, which commute  with the superintegrable Hamiltonians
\[
\{H_m,u_j\}=0\,, \qquad m=1,\ldots,\ell\,,\qquad j=1,\ldots,g\,,
\]
associated with dim $|D|=n-g$ nonregular differentials in (\ref{ab-eq}). These first integrals do not commute with $n-\ell$ first integrals  $h_{\ell+1},\ldots,h_n$. Thus, for these superintegrable systems, we can describe terms in the Riemann-Roch theorem in the following way:
\begin{itemize}
\item $n=\mathrm{deg}(D)$ - number of Abel's quadratures;
\item $g$ - number of regular differentials in Abel's quadratures;
\item dim$|D|=n-g$ - number of non-regular differentials  in Abel's quadratures;
 \end{itemize}
 and, simultaneously
 \begin{itemize}
\item $2n$ - dimension of symplectic manifold (phase space);
\item $g$ - topological genus of algebraic curve $C$ (spectral curve);
\item $n=\mathrm{deg}(D)$ - number of commuting first integrals $h_1,\ldots,h_n$;
\item $k=g$ - number of noncommuting with $h_i$ first integrals $u_1,\ldots,u_{g}$.
\item $\ell=\mathrm{dim}|D|$ - number of superintegrable Hamiltonians $H_1,\ldots,H_\ell$ commuting simultaneously with $h_i$ and $u_j$.
 \end{itemize}
 Construction of the first integrals $u_1,\ldots,u_{g}$ is a quite trivial. According to the Abel and Riemann-Roch theorems there is a unique divisor $\tilde{D}$ of minimal degree $g$ in a class of equivalent divisors which can be identified with elements of Jacobi variety $\rho(D)\in Jac(C)$ using Abel-Jacobi map. Coordinates of this divisor are the desired integrals of motion $u_1,\ldots,u_{g}$.

As an example, we consider Abel's quadratures for the inhomogeneous classical $GL(N)$ magnetic chain \cite{gek95,skl92,skl95,skott94}. The phase space  is a direct product of $M$ orbits of coadjoint action of the Lie group $GL(N)$ in fundamental vector representation, and the corresponding algebraic curve $C$ is a spectral curve of the Lax matrix defined by an irreducible equation of the form
 \begin{equation}\label{g-curve}
 C:\qquad f(x,y)=y^N+t_1(x)y^{N-1}+t_2(x)y^{N-2}+\cdots+t_{N-1}(x)y+d(x)=0\,.
 \end{equation}
In case of generic orbits polynomial $t_i(x)$ having power $iM$ in $x$
 \begin{equation} \label{hij}
 t_i(x)=a_ix^{iM}+h_{i,iM-1}x^{iM-1}+\cdots+h_{i,0}\,,\qquad a_i\in\mathbb R\,,
 \end{equation}
 contributes $iM$ functionally independent Hamiltonians $h_{i,j}$, which commute to each other with respect to the Lie-Poisson bracket which central functions are contained in polynomial $d(x)$.

Thus, we have an integrable system with $n$ first integrals in involution to each other
 \[n=\frac{MN(N-1)}{2}\,,\]
 which is integrable by Abel's quadratures (\ref{ab-eq}) on the nonhyperelliptic curve $C$ (\ref{g-curve}) of genus
\[
g=\dfrac{(N-1)(MN-2)}{2}\,.
\]
The corresponding variables of separation \cite{gek95,skl92,skl95,skott94} form a generic positive divisor $D$ with deg$(D)=n>g$. According to the Riemann-Roch theorem dimension of the linear system of equivalent divisors $|D|$ is equal to
\[
 \mathrm{dim}|D|=n-g=N - 1\,.
\]
The corresponding Abel's quadratures (\ref{ab-eq}) involve $g$ independent regular differentials
 \[
w_{i,j}=\frac{\partial f}{\partial h_{i,j}}\,\frac{dx}{\partial f/\partial y}\,,\quad\mbox{where}\quad
 \mathrm{deg}\left( \frac{\partial f}{\partial h_{i,j}}\right)\leq d-3\,,\quad\mbox{at}\quad j<iM-1\,,
\]
and dim$|D|=\ell$ non-regular differentials associated with superintegrable Hamiltonians
\begin{equation}\label{sup-h}
H_{\ell}=h_{\ell,\ell M-1}\,,\qquad \ell=1,\ldots,N-1.
\end{equation}
These superintegrable Hamiltonians commute simultaneously with $n$ integrals of motion $h_{i,j}$
  \[
  \{H_m,h_{i,j}\}=0\,,\qquad m=1,\ldots,\ell \,,\quad i=1,\ldots,N-1,\quad j=0,\ldots, iM\,,
  \]
  and with $g$ coordinates of the unique  reduced divisor $\tilde{D}$, deg$\tilde{D}=g$,
  \[\{H_m,\tilde{x}_j\}=\{H_m,\tilde{y}_j\}=0\,,
  \]
  which can be identified with elements of Jacobi variety  $\tilde{D}=\rho(D)\in Jac(C)$, see projective construction of Jacobian proposed by Chow \cite{chow}.  Here $\tilde{D}=\tilde{P}_1+\cdots \tilde{P}_g$, and $(\tilde{x}_j,\tilde{y}_j)$ are abscissa and ordinates of the point $\tilde{P}_j$ on $C$. Below instead of affine coordinates of $\tilde{D}$ we will use Mumford's coordinates.

In the present note, we prove these statements for the case $N=3$, when $dim|D|=2$. The $GL(3)$ classical magnetic chain is chosen as a sample toy-model following Sklyanin \cite{skl92}. Our calculations are quite elementary and do not involve the sophisticated algebro-geometric technique of Chow \cite{chow} or Marsden-Weinstein technique \cite{ahh93}. We only repeat original Abel's calculations \cite{ab}, because our  aim is to discuss the implementation of the standard Abel's reduction algorithm to Hamiltonian systems admitting a Lax representation with spectral parameter.

\subsection{Integrals of Abel's differential equations}
The Euler main idea \cite{eul} is that existence of algebraic trajectories is intimately connected
with the existence of a sufficient number of algebraic integrals of motion. According to \cite{eul,jac32} algebraic integrals of Abel's differential equations are suitable for finding algebraic restrictions for the form of trajectories. As a result, integrals of Abel's differential equations
 \[
 \mbox{Tr}\, \omega_j(P_1+\ldots +P_n)=\omega_j(P_1)+\cdots+\omega_j(P_n)=0\,,\qquad j=1,\ldots,g,
 \]
 have been studied in the works of Euler \cite{eul}, Abel \cite{ab}, Jacobi \cite{jac32} and Weierstrass \cite{w}, see textbooks \cite{bak,grif04}. Here $\omega_j$ are differential of the first kind on $C$, and points $P_i$ form an intersection divisor of algebraic curve $C$ with a family of auxiliary curves.

If $g$ differentials in (\ref{ab-eq}) form a basis in a space of differential forms of the first kind
\[
\omega_j=p_j(x,y)\,\frac{dx}{ f_y(x,y)}\,,\qquad \mathrm{deg}(p_j)\leq d-3\,,
\]
then
\[
\sum_{i=1}^n\int^{P_i}\omega_j+\sum_{i=1}^g\int^{\tilde{P}_i}\omega_j=const\,,\qquad j=1,\ldots,g.
\]
according to Abel's theorem \cite{ab}. Here points $\tilde{P}_i$ determine a support of the unique reduced divisor
\[\tilde{D}=\tilde{P}_1+\cdots+\tilde{P}_g\,.
\]
Coordinates of divisor $\tilde{D}$ are the so-called Abel's integrals of equations (\ref{ab-eq}) with respect to
evolution associated with dim$ |D|$ nonregular differentials \cite{bak}. These coordinates are algebraic functions
on coordinates of divisor $D$, the corresponding mechanical superintegrable systems on hyperelliptic curves are discussed in \cite{ts08,ts19k,ts20}.

In \cite{jac32} Jacobi proposed to use polynomials $\tilde{U}(x)$ and $\tilde{V}(x)$
\[
\tilde{U}(x)=\prod (x-\tilde{x}_i)=x^g+\tilde{u}_{1}x^{g-1}+\cdots+\tilde{u}_g\,,\qquad
\tilde{V}(\tilde{x}_i)=\tilde{y}_i\,,
\]
as generating functions of Abel's integrals. Therefore, Mumford remarks in \cite{mum} that these polynomials arise in the work of Jacobi only as a representation for divisors $\tilde{D}$ of degree deg$\tilde{D}=g$. In modern terms $\tilde{U}(x)$ and $y-\tilde{V}(x)$ are the so-called Mumford's coordinates of arbitrary semi-reduced divisors \cite{mum}. Other generating functions of Abel's integral, i.e. other coordinates of divisors, were proposed by Weierstrass \cite{w}.

 In the framework of a new method of integration, which was proposed by Jacobi in \cite{jac32} as a generalization of Euler's chasing of the algebraic trajectories \cite{eul}, Abel's equations (\ref{ab-eq}) determine the evolution of  integrable system with commuting first integrals  \[
h_1,\ldots, h_g;\quad H_{g+1},\ldots, H_n
\]
associated with $g$ regular and $dim |D|=n-g$ nonregular differentials on $C$. Simultaneously equations (\ref{ab-eq}) determine evolution of the second integrable system with commuting another commuting first integrals
\[
\tilde{u}_1,\ldots, \tilde{u}_g;\quad H_{g+1},\ldots, H_n\,.
\]
Solving $n+g$ algebraic equations
\[h_i=\alpha_i,\qquad H_{g+m}=\alpha_m\,,\qquad \tilde{u}_j=\tilde{\alpha}_j\]
 and dim$|D|=n-g$ differential equations for $2n$ variables we find trajectories and their different parameterizations by time for both of these systems.

In \cite{ahh93} authors propose to consider Hamiltonians
\[
\tilde{u}_1,\ldots, \tilde{u}_g;\quad H_{g+1},\ldots, H_n\,.
\]
as variables of separation which allows us to separate solution of the Jacobi's inversion problem for $g$ equations
\[
\sum_{i=1}^g\int^{\tilde{P}_i}\omega_j=const\,,\qquad j=1,\ldots,g
\]
 on a Jacoby variety $Jac(C)$ from solution of the dim$|D|=n-g$ equations associated with nonregular differentials, see discussion in textbook \cite{bt03}. Of course, it is different from the classical Euler-Jacobi construction.

\section{Abel's reduction of divisors for $GL(N)$ magnetic chain}
\setcounter{equation}{0}
The phase space of this model is a direct product of $M$ orbits of coadjoint action of the Lie group $GL(N)$ on $gl^*(N)$. Variables on this phase space $S^{(m)}_{\alpha\beta}$ satisfy to the standard  Lie-Poisson brackets
\begin{equation}\label{poi}
\left\{S^{(m)}_{\alpha_1\beta_1},S^{(k)}_{\alpha_2\beta_2}\right\}=\left(S^{(m)}_{\alpha_1\beta_1}\delta_{\alpha_2\beta_1}-S^{(m)}_{\alpha_2\beta_1}\delta_{\alpha_1\beta_2}\right)\delta_{m k}\,.
\end{equation}
Here $m,k=1,\ldots M$, $\alpha,\beta=1,\ldots, N$ and $\sum_{\alpha=1}^N S^{(m)}_{\alpha\alpha}=0$\,. Following Sklyanin \cite{skl92,skl95} we study only the generic orbits in fundamental vector representation without introducing any specific representations of algebra $gl^*(N)$.

Let $Z$ be an invertible $N\times N$ number matrix having $N$ distinct eigenvalues, $\delta_1,\ldots,\delta_M$ be some fixed numbers, and $x$ be a complex parameter (spectral parameter). In order to get Abel's equations (\ref{ab-eq}) we start with the Lax matrix
\begin{equation}\label{lax}
L(x)=Z\left(x-\delta_M+S^{(M)}\right)\cdots\left(x-\delta_2+S^{(2)}\right)\left(x-\delta_1+S^{(1)}\right)\,,
\end{equation}
and its spectral curve $C$ on the $(x,y)$ projective plane
\[
C:\qquad \det(L(x)- y)=0\,,
\]
 which has the form (\ref{g-curve})
 \[
 C:\qquad f(x,y)=y^N+t_1(x)y^{N-1}+t_2(x)y^{N-2}+\cdots+t_{N-1}(x)y+d(x)=0\,.
 \]
Using this Lax matrix we can encode original Lie-Poisson brackets (\ref{poi}) into
standard $r$-matrix form
\[
  \{\one L(u),\two L(v)\}=[r_{12}(u-v),\one L(u)\two L(v)]\,,\qquad r_{12}(u)=\frac{\rho }{u}\,\mathcal P\,,
\]
where $\mathcal P$ is the permutation operator in $\mathbb C^N\otimes \mathbb C^N$, see \cite{skl92,skl95} for details.

Separation of variables in the classical $GL(N)$ magnetic chain and the corresponding unreduced divisor $D$ of degree deg$(D)=n>g$ with dim$|D|=N-1$ are discussed in \cite{bt03,gek95, skl92,skl95, skott94}. Variables of separation $(x_i,y_i)=P_i$ form a typical positive divisor without matching points \[D=P_1+\cdots+P_n\]
having the following Mumford's coordinates
\[D=\Bigl(U(x),y-V(x)\Bigr),\]
 where
\[U(x)=\prod_{i=1}^n (x-x_i)=x^n+u_1x^{n-1}+\cdots+u_n\,,\qquad x_i\neq\pm x_j\,,\]
and
\[
V(x)=\frac{P(x)}{Q(x)}\quad \mathrm{mod}\,  U(x)\,,\qquad V(x_i)=y_i\,.
\]
It means that
\begin{equation}\label{rx}
V(x)=\frac{P(x)}{Q(x)}+R(x) U(x)\,,\qquad V(x_i)=y_i\,.
\end{equation}
Here $P(x)$ and $Q(x)$ are polynomials in $x$ which guarantee that $V(x_i)=y_i$, and $R(x)$ is an arbitrary  rational function on $C$
such that poles of $R(x)$ and zeroes of $U(x)$ are different.

According to Abel \cite{ab}, we have to consider the intersection of spectral curve $C$ with a family of curves
$y=V(x)$ and identify unreduced divisor $D$ with some part of the intersection divisor
\begin{equation}\label{int-div}
D+D'+D_\infty=0\,,
\end{equation}
Here $D_\infty$ is a suitable linear combination of points at infinity, and $V(x)$ is an interpolation function through points of both divisors $D$ and $D_\infty$.

Abscissas of points of the intersection divisor (\ref{int-div}) are zeroes of Abel's polynomial $\psi x$, which is the numerator of function
\[
f\Bigl(x,V(x)\Bigr)=V(x)^N+t_1(x)V(x)^{N-1}+t_2(x)V(x)^{N-2}+\cdots+t_N(x)=0\,,
\]
so that
\begin{equation}\label{ab-pol}
\psi x=\theta \prod_{i=1}^{n} (x-x_i) \prod_{j=1}^{m}(x-x'_j)\,,
\end{equation}
where we have to pick a function $R(x)$ in such a way that $n>m$ \cite{ab}. As a result, we obtain divisor
\[
D'=P'_1+\cdots+P'_m,\qquad n>m\]
with coordinates $\Bigl(U'(x),y-V'(x)\Bigr)$, where
\[
U'(x)=\prod_{j=1}^{m}(x-x'_j)=\mbox{MakeMonic}\left(\frac{\psi x}{U(x)}\right)\,,\qquad V'(x)=-V(x)\,\,\mbox{mod}\,\, U '(x)\,.
\]
Here the MakeMonic means that we divide the polynomial by its leading coefficient and $-$ is an inversion on $C$.

If $m>g$, we have to repeat Abel's calculations in order to get divisor $\tilde{D}$ of degree $g$ which can be identified with the representative $\rho(D)$ of a given unreduced divisor $D$ in Jacobian $Jac(C)$ \cite{chow}.

The main problem in Abel's calculations is a choice of the suitable unknown rational function $R(x)$ in (\ref{rx}). For the hyperelliptic curve, this problem was solved by Cantor's algorithm \cite{hand06}. In the next section, we solve this problem for the nonhyperelliptic spectral curve  $C$  (\ref{g-curve}).

\subsection{Reduction of divisors for $GL(3)$ magnetic chain}
According to Sklyanin \cite{skl92,skl95} separation coordinates $x_j$ are defined as poles of the vector Baker-Akhiezer function
$\Psi(x)$, which is defined as an eigenvector of the $N\times N$ Lax matrix $L(x)$
\begin{equation}\label{spectr}
L(x)\Psi(x)=y(x)\Psi(x)\,.
\end{equation}
with a suitable normalization $\bigl(\vec{a},\Psi(x)\bigr)=1$. In this case poles of $\Psi(x,y)$  are the common zeros of the vector equation
\[
\vec{a} \cdot (L(x)-y)^\wedge=0\,.
\]
where the wedge denotes the adjoint matrix \cite{bt03, kuz2,skl95}.  If we determine a  set of matrices $L^{(p)}$
\[
L^{(p)}= L\left({\rm tr}\,L^{(p-1)}\right) - (p-1)\, L^{(p-1)} L\,.
\]
with $L^{(1)}\equiv L$ and  introduce the matrix
\begin{equation}\label{b-mat}
{\cal B}(x)=
\left(\begin{array}{c}
        \vec{a}\cdot L^{(1)}(x)\;L^{-1}(x)\\
         \vec{a}\cdot L^{(2)}(x)\;L^{-1}(x)\\
      \dfrac12\;\vec{a}\cdot L^{(3)}(x)\;L^{-1}(x)\\
       \cdots\\
      \dfrac{1}{(N-1)!}\;\vec{a}\cdot L^{(N)}(x)\;L^{-1}(x)\\
\end{array}\right)\,.
\end{equation}
so that
\[
\vec{a}\cdot(L(x)-y)^\wedge
\equiv\left((-y)^{N-1},(-y)^{N-2},\ldots,1\right)\cdot\,{\cal B}(x)=0,
\]
then first Mumford's coordinate of divisor of poles $D$ is
\begin{equation}\label{U-coord}
U(x)=\mbox{MakeMonic}\,\det {\cal B}(x)\,.
\end{equation}
The finite number of equivalent by mod $U(x)$ polynomials
\begin{equation}\label{V-coord}
V_{ij}(x)=\frac{({\cal B}^\wedge(x))_{j,i}}{({\cal B}^\wedge(x))_{j,i+1}}\,,
\qquad i,j=1,\ldots,N\,,
\end{equation}
 determine second Mumford's coordinate of divisor $D$.

For $GL(3)$ magnet the simplest choice of normalization  is $\vec{a}=(0,0,1)$ and separation coordinates $x_j$ are zeroes of polynomial
\[
B(x)=L_{31}\left|
       \begin{array}{cc}
        L_{11} & L_{12} \\
        L_{31} & L_{32} \\
       \end{array}
      \right|
   +L_{32}\left|
              \begin{array}{cc}
               L_{21} & L_{22} \\
               L_{31} & L_{32} \\
              \end{array}
   \right|\,,
\]
whereas canonically conjugated momenta $p_j = \ln y_j$ are defined by
\[
y_j=A(x_j)\,,\qquad A(x)=\frac{\left|
       \begin{array}{cc}
        L_{11} & L_{12} \\
        L_{31} & L_{32} \\
       \end{array}
      \right|}{L_{32}}\,\,\mbox{mod}\,\, B(x)\,.
\]
Using a similarity transformation $WL(x)W^{-1}$  we can transform number matrix $Z$ to low-diagonal form
\[
Z=\left(
  \begin{array}{ccc}
   z_{11} & 0 & 0 \\
   z_{21} & z_{22} & 0 \\
   z_{31} & z_{32} & z_{33} \\
  \end{array}
 \right)\,,
\]
which guarantees that polynomial
\[
U(x)=\mbox{MakeMonic}\,B(x)=x^{n}+u_{1}x^{n-1}+\cdots+u_g
\]
and rational approximating function
\begin{equation}\label{v-par}
V(x)=\frac{\left|
       \begin{array}{cc}
        L_{11} & L_{12} \\
        L_{31} & L_{32} \\
       \end{array}
      \right|}{L_{32}}
\end{equation}
can be considered as Mumford's coordinates of the typical positive unreduced divisor
\begin{equation}\label{input-a}
D=\Bigl(U(x),y-V(x)\Bigr)\,,\qquad \mathrm{deg}(D)=n\,,\qquad \mathrm{dim}|D|=2\,.
\end{equation}
This divisor is an input of the Abel's reduction algorithm.

The Lax matrix $L(x)$ and normalization $\vec{a}$ generate a finite set of second Mumford's coordinates (\ref{V-coord}) for divisor of poles. To find suitable for the Abel's algorithm coordinate (\ref{v-par})  we search through this finite set of rational functions $V_{ij}(x)$ (\ref{V-coord}).

\vskip0.2truecm
\par\noindent
\textbf{First round of the reduction algorithm}\\
Substituting $y=V(x)$  into equation $f(x,y)=0$ we obtain Abel's polynomial $\psi x$ (\ref{ab-pol}) and Mumford's coordinates of divisor
\[
D'=\Bigl(U'(x),y-V'(x)\Bigr)\,,\qquad \mathrm{deg}(D')=n-1\,,\quad \mathrm{dim}|D'|=1\,,
\]
where
\[
U'(x)=\mbox{MakeMonic} \left(
L_{32}\left|
    \begin{array}{cc}
     L_{12} & L_{13} \\
     L_{32} & L_{33} \\
    \end{array}
   \right|
- L_{12}\left|
    \begin{array}{cc}
     L_{11} & L_{12} \\
     L_{31} & L_{32} \\
    \end{array}
   \right|
\right)=x^{n-1}+u'_{1}x^{n}+\cdots+u'_{n+1}
\]
and
\begin{equation}\label{vp-par}
V'(x)=\frac{\left|
    \begin{array}{cc}
     L_{12} & L_{13} \\
     L_{32} & L_{33} \\
    \end{array}
   \right|}{L_{12}}\,\,\,.
\end{equation}
Because $\mathrm{dim}|D'|=1$  we have to repeat Abel's calculations starting with the Mumford's coordinates of divisor $D'$.

To find suitable for the second round of Abel's algorithm coordinate (\ref{vp-par})  we again search through the finite set of rational functions $V'_{ij}(x)$ associated with Lax matrix and numerical normalization $\vec{a}'$ associated with $D'$.
\vskip0.2truecm
\par\noindent
\textbf{Second round of the reduction algorithm}\\
Substituting $y=V'(x)$  into equation $f(x,y)=0$ we obtain Abel's polynomial (\ref{ab-pol}) and Mumford's coordinates of divisor
\[
 D''=\Bigl(U''(x),y-V''(x)\Bigr)\,,\qquad \mathrm{deg}(D'')=g\,,\quad \mathrm{dim}|D''|=0\,,
\]
where
\begin{equation}\label{upp}
U''(x)=\mbox{MakeMonic} \left(
-L_{12}\left|
    \begin{array}{cc}
     L_{12} & L_{13} \\
     L_{22} & L_{23} \\
    \end{array}
   \right|
- L_{13}\left|
    \begin{array}{cc}
     L_{12} & L_{13} \\
     L_{32} & L_{33} \\
    \end{array}
   \right|
\right)=x^g+u''_{1}x^{g-1}+\cdots+u''_g
\end{equation}
and
\[V''(x)=-\frac{\left|
    \begin{array}{cc}
     L_{12} & L_{13} \\
     L_{22} & L_{23} \\
    \end{array}
   \right|}{L_{13}}\,.\]
It is the output of Abel's reduction algorithm for the $GL(3)$ magnetic chain.

Thus, we replace the search of a suitable rational function $R(x)$ (\ref{rx}) in a standard Abel's calculation to the iteration over a finite set of second Mumford's coordinates generated by the Lax matrix.
 \par\noindent
\textbf{End of the reduction algorithm.}
\par\noindent
 Using projective construction of the Jacobian \cite{chow} now we have to identify divisor $D''$ with representative $\rho_n(D)$ of unreduced divisor $D$ in $Jac(C)$. We can get the same result by using arbitrary numerical normalization $a$ of the vector Baker-Akhiezer function.

For the Gaudin magnets, Abel's reduction of the divisor is equivalent to the reduction of generic orbits of the corresponding loop algebra \cite{ahh93}. In a quantum case, a counterpart of this reduction of divisors is discussed in \cite{mm18,mm20}.

\subsection{Properties of divisors for $GL(3)$ magnetic chain}
Using Abel's reduction we obtained a chain of divisors
\[D\to D'\to D'' \sim\rho(D)\in Jac(C)\,,\]
where $\rho$ is the Abel-Jacobi map. Now we want to study the properties of these divisors. It is easy to calculate Poisson brackets between Mumford's coordinates of divisors
\[
\{U(x),U(\tilde{x})\}=\{U'(x),U'(\tilde{x})\}=\{U''(x),U''(\tilde{x})\}=0\,,
\]
\[
\{V(x),V(\tilde{x})\}=\{V'(x),V'(\tilde{x})\}=\{V''(x),V''(\tilde{x})\}=0\,,
\]
and
\[\begin{array}{rcl}
\{V(x),U(\tilde{x})\}&=&\phantom{-}\dfrac{1}{x-\tilde{x}}\left( V(x) U(\tilde{x}) -U(x)V(\tilde{x} )\,\dfrac{L_{32}(\tilde{x})^2}{L_{32}(x)^2}\right)\,,\\ \\
\{V'(x),U'(\tilde{x})\}&=&-\dfrac{1}{x-\tilde{x}}\Bigl( V'(x)U'(\tilde{x}) -U'(x)V'(\tilde{x} )\Bigr)\,,\\ \\
\{V''(x),U''(\tilde{x})\}&=&\phantom{-}\dfrac{1}{x-\tilde{x}}\Bigl( V''(x)U''(\tilde{x}) -U''(x)V''(\tilde{x} )\Bigr)\,,
\end{array}
\]
from which the wanted Poisson brackets for variables of separation are derived immediately using Sklyanin's method \cite{skl92,skl95}. Here $x$ and $\tilde{x}$ are spectral parameters.

The Poisson brackets between superintegrable Hamiltonians (\ref{sup-h})
 \[
H_{g+1}=h_{1,M-1}\qquad\mbox{and}\qquad H_{g+2}=h_{2,2M-1}
\]
and Mumford's coordinates of divisors $D'$ and $D''$ are equal to zero according to Abel's theorem.
Using Poisson brackets (\ref{poi}) it is easy to directly verify this fact
\[
\{H', U'(x)\}=0\,,\quad \{H', V'(x)\}=0\,,\quad\mbox{where}\quad H'=z_{11}H_{g+1}+H_{g+2}\,,
\]
and
\[
\{H_{g+1}, U''(x)\}=\{H_{g+2}, U''(x)\}=0\,,\qquad \{H_{g+1}, V''(x)\}=\{H_{g+2}, V''(x)\}=0\,.
\]
Thus, the $GL(3)$ magnetic chain is a superintegrable system with $n$ first integrals $h_{ij}$ (\ref{hij}) in involution and $g$ first integrals $u''_j$ (\ref{upp}) commuting  only with dim$|D|=2$ superintegrable Hamiltonians $H_{g+1}$ and $H_{g+2}$. First integrals $h_{ij}$ are polynomials in variables  $S^{(m)}_{\alpha\beta}$ , whereas first integrals $u''_j$ are rational functions.

\begin{prop}
Spectral curve $C$ (\ref{g-curve}) of the Lax matrix $L$ (\ref{lax}) contributes $n$ functionally independent Hamiltonians $h_{k,j}$ (\ref{hij}), which commute to each other with respect to the Lie-Poisson bracket (\ref{poi})
\[
h_{1,1},\ldots,H_{g+1}=h_{1,M-1},\,h_{2,1},\ldots,H_{g+2}=h_{2,2M-1}\,.
\]
Divisors of poles $D$ of its Baker-Akhiezer function $\Psi(x)$ (\ref{spectr}) after reduction
\[
D\to D'\to D''\sim\rho(D)
\]
can be associated with  $dim|D|=2$ families of $n$ functionally independent Hamiltonians which also commute to each other
\[
u'_1,\ldots,u'_{g+1},\,H'\,,\qquad \mbox{and}\qquad
u''_1,\ldots, u''_g;\quad H_{g+1},H_{g+2}\,.
\]
\end{prop}
According to Euler and Jacobi, we can use coordinates of divisors $D'$ and $D''$ for algebraic definition of forms of trajectories.
Following \cite{ahh93} we can also use coordinates of reduced divisor $D''$ for reduction of $n=g+2$ Abel's equations (\ref{ab-eq}) to Jacobi's canonical inversion problem on $Jac(C)$ of genus $g$ curve $C$.

\section{Conclusion}
In \cite{ts19r,ts19s,ts19k,ts19ar,ts20} we considered superintegrable systems associated with elliptic and hyperelliptic curves for which reduction of divisors is a well-studied standard part of modern cryptography on algebraic curves \cite{hand06}. In this note, we use Lax matrices and classical $r$-matrix theory to perform the reduction of divisors and to study properties of the obtained divisors on nonhyperelliptic curves.

After quite trivial calculations involving a Lax matrix, we obtain chain of divisors $D\to D'\to D''$ of degree $n$, $n-1$ and $n-2=g$ on the nonhyperelliptic spectral curve, the corresponding superintegrable Hamiltonians $H'$ or $H_{g+1},H_{g+2}$ and noncommutative integrals $u'_j$ or $u''_j$. It allows us to apply the classical Euler-Jacobi method to solve Abel's equations. Moreover, these divisors and superintegrable Hamiltonians determine three different systems of canonical coordinates, which are variables of separation for the classical $GL(3)$ magnetic chain. Substituting these different canonical variables into new separation relations we can get various integrable systems having integrals of motion of third, fourth, and sixth-order in original variables, see examples in \cite{ts15a,ts15b,ts15c,ts17c}.

Transformation of divisors $D\to D'$ is equivalent to transformation of normalization vector $\vec{a}\to\vec{a}'$. It allows us
rewrite the magic Sklyanin's  recipe
 "\textit{Take the poles of the properly normalized Baker-Akhiezer function and
the corresponding eigenvalues of the Lax operator and you obtain variables of separation}\
in the following form
 "\textit{Take the poles of the  arbitrary normalized Baker-Akhiezer function and
the corresponding eigenvalues of the Lax operator and after reduction of divisor you obtain variables of separation}\,.
We have not an algorithm for a choice of the  suitable unknown normalization, but we have  Abel's algorithms for reduction of divisors \cite{ab}.  In generic case this algorithms also involve unknown rational function $R(x)$.

In our mathematical experiment, we proved that the Lax matrix for $SL(3)$ magnetic chain contains information that is necessary for the implementation of Abel's reduction algorithm. We replace  the search of unknown function $R(x)$ on the  search through
the  finite set of second Mumford's coordinates generated by the Lax matrix. It is the main result of this note. In the forthcoming publication we apply the same idea to the reduction of divisors associated with the Lax representation for the Kowalevski top.

\section{Data Availability Statement}
The data that supports the findings of this study are available within the article.

\section{Acknowledgment}
The work was supported by the Russian Science Foundation (project 18-11-00032).


\begin{thebibliography}{10}
\bibitem{ab}
 Abel N. H.,
 \newblock{\em M\'{e}moire sure une propri\'{e}t\'{e}
g\'{e}n\'{e}rale d'une classe tr\`{e}s \'{e}ntendue de fonctions transcendantes},
Oeuvres compl\'{e}tes, Tome I, Grondahl Son, Christiania, 1881, pages 145-211,
 available from \url{ http://archive.org/details/OEuvresCompletesDeNielsHenrikAbel1881\_12/page/n167}.

\bibitem{ahh93}
Adams M.R., Harnad J., Hurtubise J.,
\newblock{\em Darboux coordinates and Liouville-Arnold integration in loop algebras},
Commun. Math. Phys., v.155, pp.385-413, 1993.

\bibitem{bt03}
Babelon O., Bernard D., Talon M.,
 \newblock{ Introduction to classical integrable systems}, Cambridge University Press, Cambridge, 2003.

\bibitem{bak}
Baker H. F.,
\newblock{Abel's theorem and the allied theory of theta functions}, Cambridge University Press, Cambridge, 1897.

\bibitem{bal03}
Ballesteros \'{A}., Musso F., Ragnisco O.,
 \newblock{\em Maximally superintegrable Gaudin magnet: A unified approach},
 Theor. Math. Phys., v.137, pp.1645-1651, 2003.


\bibitem{chow}
Chow W.L.,
\newblock{\em The Jacobian variety of an algebraic curve}, American Journal of Mathematics,
v. 76, p. 453-476, 1954.

 \bibitem{eul}
Euler L.,
\newblock{\em Probleme un corps \'{e}tant attir\'{e} en raison r\'{e}ciproque
quarr\'{e}e des distances vers deux points fixes donn\'{e}s, trouver les cas
o\'{u} la courbe d\'{e}crite par ce corps sera alg\'{e}brique},
M\'{e}moires de l'academie des sciences de Berlin v.16, pp. 228-249, 1767,\\
available from \url{http://eulerarchive.maa.org/docs/originals/E337.pdf}.

\bibitem{gek95}
Gekhtman, M.I.
\newblock{\em Separation of variables in the classical $SL(N)$ magnetic chain}, Commun.Math. Phys., v.167, pp.593-605, 1995.

\bibitem{grif04}
Griffiths P., \newblock{\em The Legacy of Abel in Algebraic Geometry},
in The Legacy of Niels Henrik Abel, Ed. Laudal and Piene, pp. 179-205, Springer, Berlin-Heidelberg, 2004.


\bibitem{jac32}
Jacobi C. G. J.,
 \newblock{\em \"{U}ber eine neue Methode zur Integration der hyperelliptischen Diff\-eren\-tial\-glei\-chun\-gen und \"{u}ber die rationale Formihrer vollst\"{a}ndigen algebraischen Integralgleichungen}, J. Reine Angew. Math., v.{32}, pp.220-227, 1846,\\
available from \url{https://gallica.bnf.fr/ark:/12148/bpt6k90215d/f147.image}


\bibitem{hand06}
Handbook of Elliptic and Hyperelliptic Curve Cryptography, ed. H. Cohen and G. Frey, Chapman and Hall/CRC, (2006).

 \bibitem{kuz92}
Kuznetsov V.B.,
\newblock{\em Quadrics on real Riemannian spaces of constant curvature: Separation of variables and connection with Gaudin magnet}
J. Math. Phys. v.33, 3240-3254, 1992.

\bibitem{kuz2}
 Kuznetsov V. B., Nijhoff  F. W.,   Sklyanin E. K.,
\newblock{\em  Separation of variables for the Ruijsenaars system},
Commun. Math. Phys., v. 189, pp. 855–877, 1997.

\bibitem{mm18}
Maillet J. M., Niccoli G.,
\newblock{\em On quantum separation of variables}, Journal of Mathematical
Physics, v. 59(9), 091417, 2018.

\bibitem{mm20}
Maillet J. M., Niccoli G., Vignoli L.,
\newblock{\em On scalar products in higher rank quantum separation of variables},
Preprint arXiv:2003.04281, 2020.

\bibitem{mir}
Miranda R.,
\newblock{Algebraic Curves and Riemann Surfaces}, Graduate Studies in Mathematics, vol 5., American Mathematical Society; UK ed., 1995.

\bibitem{mum}
 Mumford D., Tata Lectures on Theta II, Birkh\"{a}user, 1984.

\bibitem{neh72}
 Nekhoroshev N.N.,
 \newblock{\em Action-angle variables and their generalizations},
Trans. Moscow Math. Soc., v.26, pp.181-198, 1972.

\bibitem{skl92}
Sklyanin E. K.,
\newblock{\em Separation of variables in the classical integrable SL(3) magnetic chain},
 Comm. Math. Phys., v.150, no. 1, 181--191, 1992.

\bibitem{skl95}
 Sklyanin E. K.,
\newblock{\em Separation of variables-new trends}, Progr. Theoret. Phys. Suppl. (118), pp. 35-60, 1995.

\bibitem{skott94}
Scott D.R.D.,
\newblock{\em Classical functional Bethe Ansatz for $SL(N)$: separation of variables for the magnetic chain}, J. Math. Phys.,
v.35, pp.5831–5843, 1994.


\bibitem{ts08}
 Tsiganov A.V.,
\newblock{\em	On maximally superintegrable systems},
Reg.Chaotic Dyn., v.13, n.3, p.178-190, 2008.

\bibitem{ts15a}
Tsiganov A. V.,
\newblock{\em Simultaneous separation for the Neumann and Chaplygin systems},
{Regular and Chaotic Dynamics}, v.20, pp.74-93, 2015.

\bibitem{ts15b}
 Tsiganov A. V.,
\newblock{\em On the Chaplygin system on the sphere with velocity dependent potential},
{ J. Geom. Phys.}, v.92, pp.94-99, 2015.

\bibitem{ts15c}
 Tsiganov A. V.,
\newblock{\em On auto and hetero B\"{a}cklund transformations for the H\'{e}non-Heiles systems},
{Phys. Letters A}, v.379, pp.2903-2907, 2015.

\bibitem{ts17c}
Tsiganov A. V.,
\newblock{\em New bi-Hamiltonian systems on the plane},
{Journal of Mathematical Physics}, v.58, 062901, 2017.

\bibitem{ts19r}
Tsiganov A.V.,
\newblock{\em Transformation of the St\"ackel matrices preserving super\-in\-teg\-ra\-bi\-lity}//
J. Math. Phys., 2019, v. 60, 042701.

\bibitem{ts19s}
Tsiganov A.V.,
\newblock{\em Elliptic curve arithmetic and superintegrable systems},
{Physica Scripta}, v.94, 085207, 2019.

\bibitem{ts19k}
Tsiganov A.V.,
\newblock{\em The Kepler problem: polynomial algebra of non-polynomial first integrals},
{Regular and Chaotic Dynamics}, v.24, pp.353-369, 2019.

\bibitem{ts19ar}
Tsiganov A.V.,
\newblock{\em Discretization and superintegrability all rolled into one},
 Nonlinearity, v.33, pp.4924-4939, 2020.

\bibitem{ts20}
Tsiganov A.V.,
\newblock{\em Superintegrable systems and Riemann-Roch theorem},
Journal of Mathematical Physics, v.61, 012701, 2020.

\bibitem{w}
Weierstrass K.,\newblock{ \em
Bemerkungen \"{u}ber die Integration der hyperelliptischen Differential-Gleichungen}, in:
Mathematische Werke I, pp.267-274, Berlin, Mayer and M\"{u}ller,
 1895.

\end{thebibliography}
\end{document}